**REVIEW**

# Iron role paradox in nerve degeneration and regeneration

## Samira Bolandghamat | Morteza Behnam-Rassouli

Department of Biology, Faculty of
Science, Ferdowsi University of
Mashhad, Mashhad, Iran

**Correspondence**
Samira Bolandghamat, Department of
Biology, Faculty of Science, Ferdowsi
University of Mashhad, Mashhad, Iran.
Email: samira.bolandghamat@mail.
um.ac.ir

**Abstract**

Iron accumulates in the neural tissue during peripheral nerve degeneration. Some studies have already been suggested that iron facilitates Wallerian degeneration (WD) events such as Schwann cell de-differentiation. On the other hand, intracellular iron levels remain elevated during nerve regeneration and gradually decrease. Iron enhances Schwann cell differentiation and axonal outgrowth. Therefore, there seems to be a paradox in the role of iron during nerve degeneration and regeneration. We explain this contradiction by suggesting that the increase in intracellular iron concentration during peripheral nerve degeneration is likely to prepare neural cells for the initiation of regeneration. Changes in iron levels are the result of changes in the expression of iron homeostasis proteins. In this review, we will first discuss the changes in the iron/iron homeostasis protein levels during peripheral nerve degeneration and regeneration and then explain how iron is related to nerve regeneration. This data may help better understand the mechanisms of peripheral nerve repair and find a solution to prevent or slow the progression of peripheral neuropathies.

**KEYWORDS**

iron homeostasis, nerve degeneration, nerve regeneration, nerve repair, peripheral nerve, peripheral neuropathy

## 1 | INTRODUCTION

Peripheral nerve regeneration is a complex time and space-dependent cellular programming (Jessen & Mirsky, 2019). Identifying the mechanisms of nerve regeneration requires an examination of all cellular molecules and signaling pathways at different times and neural tissue locations (Li et al., 2020). The techniques of genetic modification (Raivich & Makwana, 2007; Schweizer et al., 2002), gene/protein expression assays (Funakoshi et al., 1993; Jiménez et al., 2005), administration of inhibitors or inducers (Chan et al., 2003; Kilmer & Carlsen, 1984), and concentration/activity assays of biomolecules and ions (Couraud et al., 1983; Yan et al., 2010) have increased our knowledge about cellular mechanisms of nerve regeneration. However, the multifunctionality and complex interaction of the cellular molecules and signaling pathways make it difficult to predict the output (Chang et al., 2013; van Niekerk et al., 2016). As a result, there is still ongoing research on the interactions of these components. Iron is an important cofactor for many intracellular enzymes such as DNA polymerases, DNA helicases, nitrogenases, catalases, and peroxidases (Prabhakar, 2022). In addition, it is a component of mitochondrial respiratory chain proteins, which are involved in ATP production.

Yet, a free, non-protein-bound form of iron generates free radicals by interconverting between ferrous ($Fe^{2+}$) and ferric ($Fe^{3+}$) forms, which can damage cellular components (Eid et al., 2017). Cell death processes such as apoptosis, autophagy, and ferroptosis can be induced by reactive oxygen species (ROS) (Endale et al., 2023). Iron overload promotes cell apoptosis by inducing endoplasmic reticulum stress and mitochondrial dysfunction (Schulz, 2011). Ferroptosis is a form of cell death that is caused by iron, and it is characterized by intracellular iron accumulation and an increase in lipid peroxidation (Endale et al., 2023). In the physiological state, iron toxicity is prevented by iron-binding proteins such as transferrin (Tf). Iron-binding proteins participate in iron homeostasis by absorbing, re-cycling, and storing iron (Eid et al., 2017; Schulz, 2011). Iron homeostasis impairment is observed in peripheral neuropathies. Iron overload is a common symptom of various neurodegenerative disorders with peripheral neuropathies, such as neuroferritinopathy and Friedreich's ataxia (Barbeito et al., 2009; Eid et al., 2017; Schröder, 2005). On the other hand, iron deficiency is associated with restless leg syndrome and anemia-induced peripheral neuropathy (Connor et al., 2017; Kabakus et al., 2002). Iron deficiency during development results in decreased levels of myelin basic protein (MBP) and peripheral myelin protein 22 in rats, which persists even after Fe-sufficient diet replenishment (Amos-Kroohs et al., 2019). These studies indicate that iron plays an important role in the development of the peripheral nervous system and the occurrence of peripheral neuropathies. Then, the present study aims to evaluate the changes in iron homeostasis during peripheral nerve degeneration and regeneration. This data can help to better understand the role of iron in peripheral nerve regeneration and the initiation/progression of peripheral neuropathies.

## 2 | INTRACELLULAR IRON SIGNALING PATHWAYS IN SCHWANN CELLS (SCS)

Iron, either as free form (ferric ammonium citrate [FAC] only at concentrations of 0.5 and 0.65 mM) or holo-Tf (iron-bound Tf) induces an increase in cyclic adenosine monophosphate (cAMP), phosphorylated (p)- cAMP-response element binding protein (CREB), reactive oxygen species, MBP, and myelin protein zero (P0) levels in serum-deprived SCs (Figure 1) (Salis et al., 2012). The addition of either deferoxamine (an iron chelator),

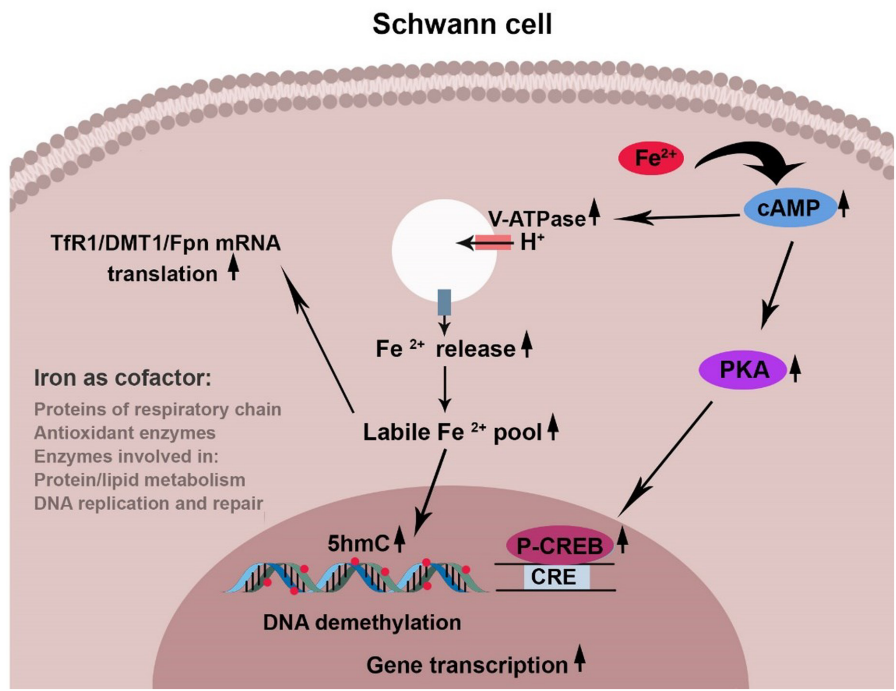

**Schwann cell**

**FIGURE 1** Intracellular iron signaling pathways in Schwann cells (SCs). Exogenous iron induces an increase in the cAMP levels in serum-deprived SCs (Salis et al., 2012). Cyclic AMP activates myelin gene transcription through two main mechanisms: the cAMP/PKA/p-CREB pathway and the production of 5-hydroxymethylcytosine (5hmC), a DNA demethylation intermediate. Cyclic AMP increases the number/function of endosomal ATPases (V-ATPases) that increase the labile iron pool required for 5hmC production (Camarena et al., 2017). After nerve injury, iron accumulates in the SCs and is associated with the up-regulation of transferrin receptor 1 (TfR1), divalent metal transporter 1 (DMT1), and ferroportin (Fpn). Iron is also a cofactor for many intracellular proteins and enzymes involved in cellular metabolism.

H-89 (a protein kinase A [PKA] antagonist), or N-acetylcysteine (a powerful antioxidant) prevents these effects of iron/hTf, indicating the role of cAMP/PKA/CREB pathway and reactive oxygen species in the pro-differentiating effect of iron (Salis et al., 2012). However, H-89 has no effect on iron/hTf-induced P0 levels in serum-deprived SCs, which means iron increases P0 expression through a PKA-independent pathway (Salis et al., 2012). These effects are not observed with iron concentrations below or above 0.5–0.65 mM (Salis et al., 2012). In the cAMP/PKA/CREB signaling pathway, cyclic AMP binds and activates PKA which then phosphorylates the transcription factor CREB (Sassone-Corsi, 2012; Shaywitz & Greenberg, 1999). The phosphorylated CREB, along with its coactivators, binds to cAMP-response elements (CREs) in the gene promoter and activates gene transcription (Chrivia et al., 1993; Kwok et al., 1994; Lundblad et al., 1995; Shaywitz & Greenberg, 1999). It was found that treatment of SCs with cAMP increases intracellular labile $Fe^{2+}$, 5-hydroxymethylcytosine (5hmC) levels, and transcription of pro-myelinating genes (Camarena et al., 2017). There is a positive correlation between 5hmC levels and gene transcription (Camarena et al., 2017; Wu & Zhang, 2017). 5-hydroxymethylcytosine is a DNA demethylation intermediate that regulates gene transcription (Wu & Zhang, 2017). It is produced by the activity of ten-eleven translocation (Tet) methylcytosine dioxygenase utilizing $Fe^{2+}$ as a cofactor (Wu & Zhang, 2017). The PKA inhibitors have no effect on the cAMP-induced increase in labile $Fe^{2+}$ and 5hmC in SCs. Administration of iron chelators or a V-ATPase inhibitor (endosomal acidification inhibitor) prevents the effects of cAMP on 5hmC, indicating the role of endosomal iron release for 5hmC generation. It appears that cAMP enhances the function/number of endosomal V-ATPases which leads to increased endosomal iron release, 5hmC generation, and gene transcription (Figure 1) (Camarena et al., 2017). In mammalian cells, iron regulates the translation of mRNAs encoding iron importer proteins (Tf receptor 1 [TfR1], divalent metal transporter 1 [DMT1]), iron-storage protein (ferritin [Fer]), and iron exporter (ferroportin [Fpn]) via iron regulatory proteins (IRP1 and IRP2) (Anderson et al., 2012). At a low intracellular iron concentration, IRP1 loses its iron–sulfur (Fe-S) cluster, binds in association with IRP2, to iron-responsive elements (IREs) in the 3′ region of TfR1 and DMT1 mRNAs, and prevents mRNA degradation by RNase, while binding of IRP1/2 to 5′ region of Fer and Fpn mRNAs prevents mRNA translation (Anderson et al., 2012; Read et al., 2021). On the other hand, at a high intracellular iron concentration, IRP1 acts as an aconitase containing the Fe-S cluster and IRP2 is degraded by the proteasome (Anderson et al., 2012; Read et al., 2021). Then, in the absence of IRP1/2 binding, TfR1 and DMT1 mRNAs are degraded, while Fer and Fpn mRNAs are translated (Anderson et al., 2012; Read et al., 2021). However, after nerve injury, iron accumulates in the SCs and coincides with the up-regulation of TfR1, DMT1, and Fpn (Martinez-Vivot et al., 2015; Raivich et al., 1991; Salis et al., 2007; Schulz, 2011). Iron is also a cofactor for many intracellular proteins and enzymes such as proteins of the mitochondrial respiratory chain (e.g., cytochrome *c*) (Cammack et al., 1990), intracellular antioxidants (peroxidases and catalase) (Cammack et al., 1990), enzymes that are responsible for DNA replication and repair (DNA polymerase, DNA helicase, DNA primase, ribonucleotide reductase, glutamine phosphoribosylpyrophosphate amidotransferase) (Cammack et al., 1990; Zhang, 2014), proteins involved in translation and post-translational modification of proteins (e.g., IRP1/2, mitochondrial aconitase, and prolyl and lysyl hydroxylases) (Cammack et al., 1990), and enzymes involved in lipid metabolism (e.g., fatty acid desaturase, stearoyl-CoA desaturase, lipoxygenase, purple acid phosphatase) (Cammack et al., 1990; Rockfield et al., 2018).

# 3 | CHANGES IN IRON HOMEOSTASIS PROTEINS AND IRON LEVELS AFTER PERIPHERAL NERVE INJURY (PNI)

Iron homeostasis proteins are expressed at very low levels in the intact nerve. After nerve injury, these proteins are up-regulated in the lesion site, distal nerve segment, and slightly in a narrow proximal segment neighboring the lesion site (Camborieux et al., 1998; Hirata et al., 2000; Madore et al., 1999; Martinez-Vivot et al., 2015; Raivich et al., 1991; Salis et al., 2007; Schulz, 2011) (Table 1). Iron hemostasis proteins that have been studied so far post-PNI, include transferrin (Tf) (Salis et al., 2007), TfR1 (Raivich et al., 1991; Salis et al., 2007), DMT1 (Martinez-Vivot et al., 2015), Fer (Rosenbluth & Wissig, 1964), Fpn (Schulz, 2011), ferroxidase ceruloplasmin (Cp) (Schulz, 2011), heme oxygenase 1 (HO-1) (Hirata et al., 2000; Hirosawa et al., 2018), and hemopexin (Hpx) (Camborieux et al., 1998; Madore et al., 1999) (Figure 2). Since the increased expression of iron homeostasis proteins occurs before macrophage invasion in Wallerian degeneration (WD), it has been suggested that their expression is early regulated by SCs and fibroblasts in the lesion site (Hirosawa et al., 2018; Madore et al., 1999). During nerve regeneration, the levels of iron homeostasis

**TABLE 1** Changes in the expression of iron homeostasis proteins and iron levels in the nerve after peripheral nerve injury in adult rodents.

| Iron hemostasis protein | Nerve injury model | Area of expression and distribution | Cell type | Time points of measurement (days post-injury [dpi]) | Expression peak (dpi) | References |
|---|---|---|---|---|---|---|
| Transferrin | Sciatic nerve crush | distal nerve segment > proximal segment | Schwann cell | 3, 5, 7, and 14 dpi (by real-time PCR and western blot analysis) | 3dpi (mRNA and protein) | (Salis et al., 2007) |
| Transferrin receptor | Sciatic nerve crush/transection | Lesion site > distal nerve segment | Schwann cell | 0, 2, 4, 6, 9, 14, and 21 dpi (Immunocytochemistry) | 4dpi | (Raivich et al., 1991) |
| | Sciatic nerve crush | Proximal segment > distal segment | Schwann cell | 3, 5, 7, and 14dpi (Western blot analysis) | 14dpi | (Salis et al., 2007) |
| | Sciatic nerve crush | Distal nerve segment | Schwann cell | 1, 7, and 14dpi (real-time PCR) | 1dpi | (Schulz, 2011) |
| | Sciatic nerve crush | Lesion site, distal nerve segment | Schwann cell | 0, 1, 2, 3, 7, 14, and 21 dpi (Immunohistochemistry) | 2dpi | (Hirata et al., 2000) |
| Divalent metal transporter 1 | Sciatic nerve crush | Distal nerve segment | Schwann cell | 1, 7 and 14dpi (real-time PCR) | N/C | (Schulz, 2011) |
| | Sciatic nerve crush | Lesion site > distal nerve segment | Schwann cell | 7, 14, 21, 35, and 56dpi by real-time PCR and (western blot analysis) | 7dpi (mRNA) 14 dpi (protein) | (Martinez-Vivot et al., 2015) |
| Ferroportin | Sciatic nerve crush | Distal nerve segment | Schwann cell | 1, 7 and 14dpi (real-time PCR)7 dpi | 7dpi | (Schulz, 2011) |
| Ceruloplasmin | Sciatic nerve crush | Distal nerve segment | Schwann cell | 1, 7, and 14dpi (real-time PCR) | N/C | (Schulz, 2011) |
| Heme oxygenase 1 | Sciatic nerve crush | Distal nerve segment | Schwann cell | 0, 3, 7, 14 and 28dpi (real-time PCR and western blot analysis) | 3dpi (mRNA and protein) | (Kim et al., 2019) |
| | Sciatic nerve crush | Lesion site, distal nerve segment | Schwann cell | 0, 1, 2, 3, 7, 14, and 21 dpi Immunohistochemistry) (and immunoblot analysis | 2dpi | (Hirata et al., 2000) |
| Hemopexin | Sciatic nerve crush/transection (axotomy) hypoglossal nerve transection (axotomy) sympathic trunk transection | Lesion site, distal nerve segment | Schwann cell | 2, 5, and 7dpi (in situ hybridization and immunohistochemistry) | 2dpi (mRNA) 7dpi (protein) | (Camborieux et al., 1998) |
| | Sciatic nerve transection (axotomy) | Proximal and distal segments | N/D | 2, 7, 15, 60, and 90dpi Immunohistochemistry) (and immunoblot analysis | 2 and 60dpi (distal segment) 2 and 15dpi (proximal segment) | (Madore et al., 1994) |
| | Sciatic nerve crush | Lesion site, distal nerve segment | N/D | 2, 7, 15, 60, and 90dpi Immunohistochemistry) and immunoblot analysis | 2dpi | (Madore et al., 1994) |
| | Sciatic nerve crush/transection (axotomy) | Distal nerve segment | N/D | 2, 7, and 14dpi (real-time PCR and immunoblotting and immunoprecipitation) | 2dpi (mRNA) 2 and 7dpi (protein) | (Madore et al., 1999) |

| Iron | Nerve injury model | Area of expression | Cell type | Time points of measurement (days post-injury (dpi)) | Accumulation peak (dpi) | References |
|---|---|---|---|---|---|---|
| Iron accumulation | Sciatic nerve crush | Lesion site > distal nerve segment | N/D | 7, 14, 21, 35, and 56dpi (atomic absorption spectroscopy) | 14-21 dpi | (Martinez-Vivot et al., 2015) |
| Increased uptake of radioactive iron | Sciatic nerve crush | Lesion site, distal nerve segment | N/D | 0, 1, 3, 6, 10, and 15 dpi (Autoradiography) | 3dpi | (Raivich et al., 1991) |

*Note:* (>) indicates more intensity of expression; dpi means days post-injury; N/C means no change; N/D means no data.

proteins are progressively decreased and return to normal levels in the intact nerve (Camborieux et al., 1998; Hirata et al., 2000; Hirosawa et al., 2018; Madore et al., 1999; Martinez-Vivot et al., 2015; Raivich et al., 1991; Rosenbluth & Wissig, 1964; Salis et al., 2007).

## 3.1 | Tf and TfR1

Tf is the main iron-transport protein in the extracellular fluid. It delivers iron to cells by binding to TfR1 on the cell surface and endocytosis (Schulz, 2011). Iron binding to Tf avoids the toxic effect of free iron in the extracellular space (Eisenstein, 2000). In the cell, the TfR1 levels are controlled by negative feedback from the intracellular iron concentration (Eisenstein, 2000); however, after PNI, this control is lost and Tf/TfR1 levels are increased in the phagocytic SCs and regenerating motor neurons (Raivich et al., 1991; Salis et al., 2007; Schulz, 2011). This event is accompanied by increased endoneurial iron uptake in the lesion site (Raivich et al., 1991). The increased Tf levels in SCs and neurons are a result of the increased gene expression and its uptake from the systemic circulation (Raivich et al., 1991; Salis et al., 2007). In axolotl regenerating axons, Tf is carried via fast anterograde transport and released from the growth cones (Kiffmeyer et al., 1991). Tf has a cytoplasmic location in SCs and axons (Lin et al., 1990). It is seen more at the nodes of Ranvier of myelinated fibers (Lin et al., 1990). Tf is more abundant in a myelinated peripheral nerve than in an unmyelinated peripheral nerve, likely due to its role in myelination (Lin et al., 1990). It has been found that ablation of TfR1 reduces embryonic SC proliferation, maturation, and postnatal axonal myelination (Santiago González et al., 2019). Moreover, the addition of iron or holo-Tf to the SCs culture prevents cell de-differentiation induced by serum withdrawal, as is evidenced by increased expression of markers of SCs differentiation such as MBP and P0 (Salis et al., 2002, 2012). Holo-Tf induces MBP and P0 expression, respectively, through cAMP/PKA/CREB-dependent and PKA-independent signaling pathways, in the serum-deprived SCs (Salis et al., 2012).

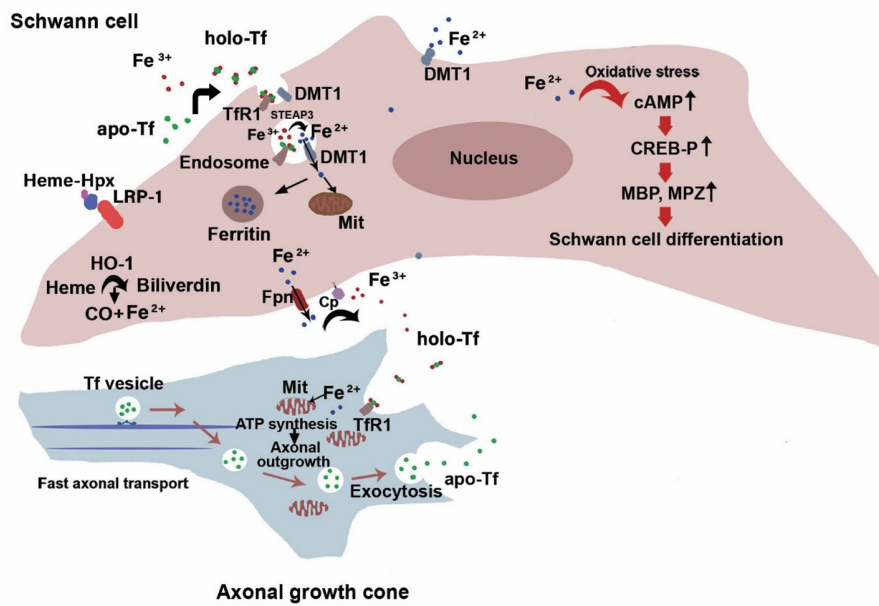

**FIGURE 2** Iron trafficking between Schwann cell and growth cone during peripheral nerve regeneration. Iron enters the Schwann cell by binding holo-transferrin (holo-Tf, iron-bound apo-transferrin[apo-Tf]) to transferrin receptor 1 (TfR1) or via divalent metal transporter 1 (DMT1) (Martinez-Vivot et al., 2015; Schulz, 2011). The endosomal enzyme, metalloreductase STEAP3 reduces $Fe^{3+}$ to $Fe^{2+}$ which is then delivered to the cytosol by DMT1 (Schulz, 2011). $Fe^{2+}$ enters mitochondria, is stored in the ferritin molecule, or accumulates intracellularly as a "labile pool" (Not shown) (Schulz, 2011). The increased levels of iron induce the expression of myelin basic protein (MBP), myelin protein zero (MPZ), and then Schwann cell differentiation through the increase in levels of cAMP and p-CREB (Salis et al., 2002, 2012; Schulz, 2011). On the other hand, the heme-hemopexin (Hxp) complex binds to low-density lipoprotein receptor-related protein (LRP-1) and enters the Schwann cell (Camborieux et al., 1998; Tolosano et al., 2010). Heme inside the Schwann cell is degraded by heme oxygenase-1 (HO-1) into biliverdin, carbon monoxide (CO), and $Fe^{2+}$ (Hirata et al., 2000). $Fe^{2+}$ leaves Schwann cell via ferroportin (Fpn) and is then converted by ceruloplasmin (Cp) into $Fe^{3+}$ for binding to apo-Tf (Schulz, 2011). Holo-Tf binds to TfR1 on the axonal growth cone (Raivich et al., 1991). In the axonal growth cone, iron enters mitochondria for the synthesis of ATP which provides the energy necessary for axonal growth (Schulz, 2011). On the other hand, the apo-Tf-containing vesicles synthesized in the neuronal cell body are transferred to the growth cone through fast axonal transport and released from the growth cone through exocytosis (Kiffmeyer et al., 1991). Apo-Tf, apo-transferrin; holo-Tf, holo-transferrin; TfR1, transferrin receptor 1; DMT1, divalent metal transporter 1; MBP, myelin basic protein; Mit, mitochondrion; MPZ, myelin protein zero; Hxp, hemopexin; LRP-1, low-density lipoprotein receptor-related protein; HO-1, heme oxygenase-1; CO, carbon monoxide; Fpn, ferroportin; Cp, ceruloplasmin.

## 3.2 | DMT1

DMT1 is responsible for the cellular uptake of non-Tf-bound iron through the cellular and endosomal membranes (Martinez-Vivot et al., 2013, 2015). In the peripheral nerve, DMT1 is localized in the plasma membrane of SCs (Martinez-Vivot et al., 2013). In a recent study, after a sciatic nerve crush injury, an increase in DMT1 mRNA and protein levels was observed at the lesion site and distal stump (Martinez-Vivot et al., 2015). However, the other study could not find any change in DMT1 mRNA in the distal nerve after PNI (Schulz, 2011). It is supposed that the DMT1 level increase is a result of activated inflammatory processes during WD as is observed in CNS (Martinez-Vivot et al., 2015; Urrutia et al., 2013). In differentiated PC12 cells (a model for neuronal differentiation into sympathetic-neuron-like cells (Hu et al., 2018)), DMT1 is responsible for the majority of iron uptake (Mwanjewe et al., 2001; Schonfeld et al., 2007). Ablation of DMT1 reduces embryonic SC proliferation, maturation, and postnatal myelination (Santiago González et al., 2019). Ablation of DMT1 also down-regulates TfR1 and vice versa (Santiago González et al., 2019). After chronic constriction injury of the sciatic nerve, the expression of DMT1 mRNA without iron-responsive element ([−] IRE mRNA) and DMT1 protein are increased in the spinal cord dorsal horn, with a peak at 7 days post-injury (Xu et al., 2019). Since the intracellular iron levels control the binding of iron regulatory proteins to the IRE and stabilization of the mRNA (Anderson et al., 2012), it could be suggested that the increased expression of (−) IRE mRNAs after PNI may be a strategy taken by cells for iron accumulation by escaping from the inhibitory effect of high iron levels on the expression of iron importer proteins.

## 3.3 | Fer

Fer is an intracellular iron-storage protein. It has ferroxidase activity, which converts ferrous into ferric iron to be deposited inside the Fer core (Santiago González et al., 2019). There are two types of Fer inside the cell: cytosolic and mitochondrial Fer (Arosio & Levi, 2010). Fer expression is controlled by intracellular iron concentration. High iron concentration increases Fer expression (Anderson et al., 2012). There are no reports of levels of nerve Fer after PNI. However, considering iron accumulation after PNI (Martinez-Vivot et al., 2015; Raivich et al., 1991), it can be thought that Fer levels are increased. Ferric ammonium citrate (FAC)-induced iron overload in differentiated PC12 cells increases the expression of Fer subunits mRNA (Helgudottir et al., 2019). Ablation of Fer reduces embryonic SC proliferation, maturation, and postnatal myelination. These defects are more severe in Fer knockout mice than in TfR1 or DMT1 knockout mice (Santiago González et al., 2019). Moreover, the neurons of the cultured spinal ganglia, similar to those of the intact ganglia, uptake exogenous Fer (Rosenbluth & Wissig, 1964).

## 3.4 | Fpn and Cp

SCs express Fpn and Cp, two proteins that partner to efflux iron from SCs (Camborieux et al., 1998; Schulz, 2011). The expression of Fpn mRNA is greater in differentiated PC12 cells than in undifferentiated cells (Helgudottir et al., 2019). Up-regulation of Fpn has been shown after sciatic nerve crush injury (Schulz, 2011). The sciatic nerve crush injury in Cp knockout mice results in impaired axonal regeneration and motor recovery (Mietto et al., 2021; Schulz, 2011). Additionally, knocking-out Cp in mature myelinating SCs reduces the expression of myelin proteins and induces oxidative stress (Santiago González et al., 2021). Knocking-out Cp also causes increased levels of TfR1, DMT1, and Fer in SCs (Mietto et al., 2021).

## 3.5 | Heme-related proteins

In phagocytic SCs, HO-1 is induced, which catalyzes the oxidation of heme to biliverdin, CO, and $Fe^{2+}$ (Hirata et al., 2000; Kim et al., 2019). HO-1 is thought to be another source of iron overload in the cells (Liao et al., 2021). It also protects cells from the toxicity of free heme (Ryter, 2021). After PNI, HO-1 up-regulates in the dorsal root ganglion (DRG) and spinal cord (Chen, Chen, et al., 2015; Liu et al., 2016). Another study revealed that, after PNI, HO-1 is expressed in the microglia of the spinal cord but not neurons and astrocytes (Liu et al., 2016). Induction of HO-1 after PNI inhibits microglia activation (Liu et al., 2016), expression of pro-inflammatory cytokines (Chen, Chen, et al., 2015), and neuropathic pain (Chen, Chen, et al., 2015; Liu et al., 2016). Hpx is a heme-scavenger protein (Tolosano et al., 2010), which is up-regulated in SCs and fibroblasts after PNI (Camborieux et al., 1998; Madore et al., 1994, 1999; Swerts et al., 1992). Chronic axotomy results in sustained elevation of Hpx levels up to 3 months after nerve injury (Madore et al., 1994). After binding to heme, the heme-Hpx complex enters the cell by binding to its receptor, low-density lipoprotein receptor-related protein (LRP-1), and endocytosis (Tolosano et al., 2010). The expression of LRP-1 has increased in SCs after PNI (Campana et al., 2006; Gaultier et al., 2008; Mantuano et al., 2008, 2015), indicating the increased uptake of heme by SCs. TNF-α can induce LRP-1 expression in SCs (Campana et al., 2006). LRP-1 plays a role in SC survival (Campana et al., 2006; Mantuano et al., 2011; Orita et al., 2013) and migration after PNI (Mantuano et al., 2008). In sum, it seems that after PNI, SCs employ all mechanisms of iron accumulation, from increased cellular uptake of iron and the heme to iron release from heme by HO-1 activity.

## 3.6 | Iron

The iron concentration in normal sciatic nerve tissue of the rat is $36.90 \pm 1.00 \mu g/g$ (Liu et al., 2022). After nerve crush injury, iron accumulation is observed at the lesion site and distal stump (Martinez-Vivot et al., 2015; Raivich et al., 1991). In a recent study, the maximum intracellular iron concentration was observed at the lesion site and distal stump 2–3 weeks after nerve crush injury, which was approximately 2.5-fold higher than the normal concentration (Martinez-Vivot et al., 2015). Another study has reported a peak of radioactive iron uptake at the lesion site 3 days after a nerve crush injury (Raivich et al., 1991). PNI also leads to increased uptake and accumulation of iron in the central nervous system (Graeber et al., 1989; Xu et al., 2019). In a recent study, iron levels were increased in the spinal cord dorsal horn following chronic constriction injury of the sciatic nerve, first observed at 3 days with a peak at 7 days post-injury (Xu et al., 2019). Furthermore, Perl's staining of the sciatic nerve explants has shown ferric ion deposition in 1 day, with a peak at 3 days postexplant (Han et al., 2022).

## 4 | ROLE OF IRON IN WD AND NERVE REGENERATION

Some authors have suggested that the up-regulation of iron homeostasis proteins and high iron levels are needed for WD events (Hirata et al., 2000; Kim et al., 2019; Salis et al., 2007). HO-1 is proposed to be involved in myelin degradation, SC de-differentiation, and SC proliferation by inducing oxidative stress (Hirata et al., 2000; Kim et al., 2019). However, it has also been suggested that HO-1 plays a role in cytoprotection against oxidative stress (Yardım et al., 2021) as the addition of Znpp, the competitive inhibitor of HO-1, reduces the viability of serum-deprived PC12 cells (Lin et al., 2010; Martin et al., 2004). In addition, the activation of the phosphatidylinositol 3-kinase (PI3K)/Akt pathway, a survival signaling pathway (Barzegar-Behrooz et al., 2022; Li et al., 2001), increases the expression of HO-1 in PC12 cells (Martin et al., 2004). On the other hand, the literature has shown that an increased level of iron homeostasis proteins and iron accumulation is observed a few days after WD initiation (Camborieux et al., 1998; Hirata et al., 2000; Madore et al., 1994, 1999; Martinez-Vivot et al., 2015; Raivich et al., 1991; Salis et al., 2007; Schulz, 2011) when SCs have de-differentiated (Hirata et al., 2000). Thus, it appears that iron is not required for the initiation of WD, SC de-differentiation, and proliferation (Santiago González et al., 2019), but rather is required for the late stage of WD and initiation of nerve regeneration. Iron polarizes macrophages from a pro-inflammatory M1 to an anti-inflammatory M2 Phenotype (Agoro et al., 2018; Chen et al., 2020). Also, induction of HO-1 inhibits microglia activation (Chen, Chen, et al., 2015) and expression of pro-inflammatory cytokines after PNI (Liu et al., 2016). Exogenous iron uptake has increased in differentiated PC12 cells compared to non-differentiated cells (Mwanjewe et al., 2001). Moreover, iron promotes SC differentiation in the culture (Salis et al., 2002, 2012). Iron increases cAMP levels and CREB phosphorylation, which induces the expression of myelin proteins (Salis et al., 2012). It has been suggested that iron accumulation is required for the initiation of myelination in the CNS, as high iron concentration is present in myelin and cytoplasm of oligodendrocytes (Connor & Menzies, 1996). Iron is a cofactor for enzymes responsible for the synthesis and degradation of myelin lipids, such as fatty acid desaturase and lipid dehydrogenases (Connor & Menzies, 1996). Also, since iron is a cofactor for many enzymes involved in protein synthesis (Pain & Dancis, 2016), high iron levels may be required for the increase in the synthesis of neurotrophic factors and their receptors by SCs that should be elucidated (Han et al., 2022). After PNI, there is an increased expression of collagen and procollagen hydroxylases (Araki et al., 2001; Chen, Cescon, et al., 2015; Chernousov et al., 1999; Isaacman-Beck et al., 2015; Siironen et al., 1992a, 1992b, 1996) requiring iron as a cofactor (Gelse et al., 2003). Collagen types IV, V, and VI are a component of SC basal lamina involved in SC migration, spreading, myelination, M2 macrophage polarization, axonal growth, and axonal guidance (Chen, Cescon, et al., 2015; Chernousov et al., 2001, 2006; Erdman et al., 2002; Fang & Zou, 2021; Isaacman-Beck et al., 2015;

Lv et al., 2017; Sun et al., 2022). SCs can proliferate and normally grow on the electrospun silk fibroin scaffolds with different concentrations of incorporated iron oxide nanoparticles (1–10 wt. % iron oxide nanoparticles) (Taneja, 2013). SC growth was better on scaffolds containing a higher concentration of incorporated iron particles (7 wt. %). Furthermore, nerve growth factor (NGF) levels were higher in the electrospun silk fibroin scaffolds containing a concentration of 3 wt. % iron oxide nanoparticles than scaffolds with no or lower concentrations of iron oxide nanoparticles (Taneja, 2013). However, NGF levels were reduced in scaffolds with a concentration of 5 wt. % iron oxide nanoparticles (Taneja, 2013). Iron up to a concentration of 500 μM has not shown any cytotoxicity against PC12 cells for up to 5–6 days (Hong et al., 2003; Kim et al., 2011). Iron nanoparticles also have any cytotoxicity against cultured Schwann cells up to a concentration of 2 μg/mL (intercellular iron concentration of $1.21 \pm 0.08$ pg/cell) for 72 h (Xia et al., 2016, 2020). Moreover, the iron solution (FeCl$_2$) with concentrations of 10, 100, or 500 mM in combination with NGF increases the viability of the serum-deprived PC12 cells (about 2-fold) compared to NGF treatment alone (Hong et al., 2003). In PC12 cells, iron causes a dose-dependent increase in the expression of p-ERK, p-Bad, and Bcl-2 (Kim & Yoo, 2013). The phosphorylated Bad and Bcl-2 are anti-apoptotic proteins that reduce the release of cytochrome c from mitochondria (Kim & Yoo, 2013; Yardım et al., 2021). Furthermore, iron enhances the viability of SCs (Han et al., 2022). In a recent study, SC viability was approximately 140% in a 2.5 mM ferric ammonium citrate solution, but it decreased with higher concentrations (Han et al., 2022). Then, it seems, the effects of iron on the cells depend on its concentration and the cellular capacity of iron chelating (Zhao et al., 2013). Iron acts as a redox sensor in the cell (Outten & Theil, 2009) and also increases the synthesis of intracellular antioxidants such as glutathione (Cozzi et al., 2013; Lall et al., 2008). Therefore, during WD, an increase in iron levels in SCs may have a protective effect. Neurite outgrowth initiation and elongation are hindered by the iron chelator addition to the DRG culture (Schulz, 2011). Previous studies have demonstrated that iron enhances neurite outgrowth in cultured PC12 cells, whether with or without the presence of NGF (Hong et al., 2003; Katebi et al., 2019; Kim et al., 2011; Sadeghi et al., 2023; Zarei et al., 2022). The effect of iron on the neurite outgrowth of PC12 cells is believed to be mediated by integrin β1 (Hong et al., 2003; Kim et al., 2011). The expression of integrin β1 in NGF-treated PC12 cells increases as a result of an increase in iron concentration in the culture (Kim et al., 2011). Studies have shown that integrin β1 is involved in SC myelination (Nodari et al., 2007; Pellegatta et al., 2013). Inhibiting the integrin β1 function prevents myelination and causes a demyelinating neuropathy with disrupted radial sorting of axons (Nodari et al., 2007; Pellegatta et al., 2013). Integrin β1-null SCs can migrate and proliferate but do not extend processes around axons (Nodari et al., 2007; Pellegatta et al., 2013). Iron enhances the NGF signaling in PC12 cells (Kim et al., 2011; Yoo et al., 2004). It increases the levels of p-ERK 1/2 in a dose-dependent manner (Kim et al., 2011; Yoo et al., 2004). Phosphorylated ERK enhances SC survival and axonal outgrowth (Hausott & Klimaschewski, 2019). Axonal growth cones have abundant mitochondria, providing ATP required for protein synthesis, cytoskeleton assembly, and axonal transport. Regarding the role of iron in ATP synthesis and the mitochondria function, it is conceivable that the growth cone has a high iron demand (Schulz, 2011). Schulz suggested that SCs deliver the iron required for the growth cone mitochondria (Schulz, 2011) (Figure 2). Knocking out the gene of Cp, an iron exporter, on Schwann cells reduces mitochondrial ferritin (a marker of mitochondrial iron content) in axons and impairs nerve regeneration following PNI (Schulz, 2011). Iron overload increases the active matrix metalloproteinase-9 (MMP-9) and MMP-1 levels in the CNS (García-Yébenes et al., 2018; Mairuae et al., 2011). Elevated levels of MMP-9 are also reported after PNI (Remacle et al., 2018; Siebert et al., 2001). Matrix metalloproteinase-9 has been implicated in macrophage recruitment, SC migration and differentiation, axonal outgrowth, and remyelination after PNI (Verslegers et al., 2013). Migration of SCs can be promoted by the Hpx domain of MMP-9 and LRP-1 (Mantuano et al., 2008), which are both up-regulated after PNI. The peripheral nerve injury increases the levels of iron homeostasis proteins and iron in the DRG and dorsal horn of the spinal cord beyond the lesion site (Chen, Chen, et al., 2015; Liu et al., 2016; Xu et al., 2019), which is likely the result of retrogradely transported signals from the lesion site (Mietto et al., 2021). As mentioned above, the majority of the studies have focused on the proteins involved in iron homeostasis, and relatively little is known about the effects of iron excess or deficiency on peripheral nerve degeneration and regeneration. Systemic or local administration of Fe$_3$O$_4$ nanoparticles after PNI improves the morphological, functional, and electrophysiological indices of the rat sciatic nerve (Chen et al., 2020; Pop et al., 2021; Tamjid et al., 2023). Intraperitoneal administration of omega-3-coated Fe$_3$O$_4$ nanoparticles, either a dosage of 10 mg/kg/day or 30 mg/kg/day for 1 week, has improved morphological and functional indices of the rat sciatic nerve after nerve crush, with greater effects observed at a dosage of 30 mg/kg (Tamjid et al., 2023). Also, the oral administration of chitosan-coated iron nanoparticles (2.5 mg/kg/day) for 21 days improves the morphological and functional indices of the sciatic nerve and slightly increases serum NGF levels after sciatic nerve compression injury (Pop et al., 2021). In a recent study, using a multilayered nerve conduit loaded with melatonin and $Fe_3O_4$ nanoparticles improved the morphological, functional, and electrophysiological indices of the rat sciatic nerve at 16 weeks postoperation (Chen et al., 2020). The multilayered nerve conduit loaded with melatonin and $Fe_3O_4$ magnetic nanoparticles induced the macrophage polarization to the M2 phenotype in the nerve (Chen et al., 2020). Moreover, the loading of conduits with melatonin and $Fe_3O_4$ nanoparticles decreased the expression of pro-inflammatory cytokines (IL-6, TNF-α, and IFNγ), neuronal nitric oxide synthase, and vimentin (a marker of fibrosis) in the nerve (Chen et al., 2020). On the other hand, it increased the expression of the anti-inflammatory cytokine IL-10, S100 (Schwann cell marker), neurofilament protein 200 (a neuronal marker), MBP, and β3-tubulin (a neuronal marker) (Chen et al., 2020). In a recent study, systemic administration of iron solution exacerbated the DRG neuronal loss caused by sciatic nerve transection, as was demonstrated by the decreased mean number of neurons and volume of DRG (Mohammadi-Abisofla et al., 2018). Following iron administration, neuronal loss in the DRG of the injured nerve has been observed (Mohammadi-Abisofla et al., 2018), while iron usually accumulates in the DRG without significant toxicity after PNI. It may be explained by a decreased iron-chelating capacity at a specific iron concentration caused by an immediate increase in intracellular iron levels after iron administration.

# 5 | CONCLUSION

After PNI, the expression of all proteins involved in iron homeostasis is increased in SCs and axons, which shows a high demand for iron during this period. Based on previous studies, iron homeostasis proteins play a role in SC differentiation, myelination, and axonal outgrowth. However, the intracellular signals inducing the expression of these proteins are yet to be clarified. On the other hand, there is little data about the effects of iron (iron deficiency/excess) on peripheral nerve regeneration, which needs further research. Moreover, the role of iron in the cellular signaling pathways involved in peripheral nerve regeneration remains to be elucidated.

## AUTHOR CONTRIBUTIONS

Conceptualization, S.B. and M.B.R.; writing-original draft preparation, S.B. and M.B.R.; writing—review and editing, S.B. and M.B.R.; supervision, S.B. and M.B.R. All authors have read and agreed to the published version of the manuscript.

## ACKNOWLEDGMENTS
Declared none.

## FUNDING INFORMATION
No funding information provided.

## CONFLICT OF INTEREST STATEMENT
The author declares no conflict of interest, financial,d or otherwise.

## DATA AVAILABILITY STATEMENT
Not applicable.

## ETHICS STATEMENT
Not applicable.

## ORCID
*Samira Bolandghamat*  https://orcid.org/0000-0002-6627-7796

---

**How to cite this article:** Bolandghamat, S., & Behnam-Rassouli, M. (2024). Iron role paradox in nerve degeneration and regeneration. *Physiological Reports*, *12*, e15908. https://doi.org/10.14814/phy2.15908